\documentclass[11pt]{article}
\usepackage{moriond,epsfig}

\def\Journal#1#2#3#4{{#1} {\bf #2}, #3 (#4)}

\def\PLB{{\em Phys. Lett.}  B}

\def\PRD{{\em Phys. Rev.} D}

\def\be{\begin{equation}}
\def\ee{\end{equation}}
\def\bea{\begin{eqnarray}}
\def\eea{\end{eqnarray}}

\begin{document}
\rightline{IPPP/03/26, DCPT/03/52}
\rightline{UG-FT-151/03}
\rightline{hep-ph/0305119}
\rightline{Presented at XXXVIII Rencontres de Moriond:}
\rightline{Electroweak Interactions and Unified Theories}
\rightline{Les Arcs, France, March 15--22, 2003}
\vspace*{1.5cm}
\title{Some consequences of brane kinetic terms for bulk fermions}

\author{F. del Aguila $^1$, M. P\'erez-Victoria $^{1,2}$ and 
J. Santiago $^{3,}$ \footnote{Talk given by this author.}}

\address{$^1$ Departamento de F{\'\i}sica Te\'orica y del Cosmos 
and Centro Andaluz de F{\'\i}sica de \\ Part{\'\i}culas Elementales 
(CAFPE), Universidad de Granada, E-18071 Granada, Spain\\
 $^2$ Dipartimento di Fisica ``G. Galilei'', Universit\`a di Padova and \\
INFN, Sezione di Padova, Via Marzolo 8, I-35131 Padua, Italy \\
 $^3$ Institute for Particle Physics Phenomenology, University of 
Durham, \\ South Road, Durham DH1 3LE, UK }

\maketitle\abstracts{
In theories with extra dimensions there are generically brane kinetic 
terms for fields living in the bulk. These modify the masses and 
wave functions of the Kaluza-Klein expansions, and then the effective 
four dimensional gauge and Yukawa couplings of the corresponding 
modes. Here, we discuss some phenomenological consequences of fermion 
brane kinetic terms, emphasizing their implications for models with 
low compactification scales, whose agreement with experiment can 
be improved, and observable effects at collider energies, 
like the production of new fermions.    
}

\section{Introduction}

Particle physics models with extra dimensions are interesting not 
only because they are motivated, for instance, by string theories, 
but also because they offer a playground for new theoretical insights 
on longstanding standard model (SM) puzzles, like the different 
mass hierarchies, as well as new physics at low energy observable 
at future colliders. These models can have fields localized 
on lower dimensional defects (branes) and fields propagating 
in all the dimensions (bulk fields). The latter acquire kinetic terms
localized on the branes through their interactions with 
the former~\cite{Dvali:2000hr} or due to an orbifold
projection~\cite{Georgi:2000ks}.
These brane kinetic terms (BKT) require renormalization and hence run
with the scale, which indicates that it is natural to take them into
account from the very beginning.  
Moreover, they can modify phenomenological predictions of these
theories
~\cite{Dvali:2000rx,Cheng:2002iz,Carena:2002me,Kyae:2002fk,delAguila:2002mt,Davoudiasl:2002ua}. 
We have studied in detail the
most general BKT for scalars, 
fermions and gauge bosons in~Ref. \cite{delAguila:2003bh}.
Banishing possible critical behaviours (which can be dealt with by
``classical renormalization''), their effects are 
typically controlled by the size of their coefficients. Therefore, one
can expect small corrections when these terms are generated
radiatively~\footnote{These small corrections can nevertheless be
observable in certain cases~\cite{Cheng:2002iz}.}. Here we allow for 
BKT of arbitrary size and describe some of their phenomenological
implications.

The phenomenology of 
BKT for gauge fields has been addressed in Ref. \cite{Carena:2002me}
(see \cite{Davoudiasl:2002ua,Carena:2002dz}for Randall-Sundrum
models).  
In the following we discuss some consequences of BKT for 
bulk fermions (see also \cite{delAguila:2002mt}). We neglect possible 
boson BKT for simplicity.
As in the bosonic case, the masses and 
wave functions of the Kaluza-Klein (KK) modes are modified, which in
turn corrects their gauge and Yukawa couplings. 

We consider a five-dimensional model with the fifth dimension $y$
compactified on an orbifold $S^1 / Z_2$ with radius $R$. 
The general free Lagrangian for a massless bulk fermion $\Psi$ reads  
(see Ref.~\cite{delAguila:2003bh} for notation; in particular 
$\delta _0 \equiv \delta (y), 
\delta _{\pi} \equiv \delta (y - \pi R)$, and sum over $I=0,\pi$ is
understood): 
\begin{equation}
\begin{array}{rcl}
{\mathcal L}
&=&
\int_{-\pi R}^{\pi R} \mathrm{d}y\;
\Big\{
\big[1+ a^L_I\delta_I 
\big]
\bar{\Psi}_L \mathrm{i} \not \! \partial \Psi_L
+
\big[1+ a^R_I\delta_I
\big]
\bar{\Psi}_R \mathrm{i} \not \! \partial \Psi_R \\
&&
\phantom{\int_{-\pi R}^{\pi R} \mathrm{d}y}
-\big[1+ b_I\delta_I
\big] \bar{\Psi}_L \partial_y \Psi_R
- b_I \delta_I \big( \partial_y \bar{\Psi}_R\big) \Psi_L \\
&&
\phantom{\int_{-\pi R}^{\pi R} \mathrm{d}y}
+\big[1+ c_I\delta_I \big] \bar{\Psi}_R \partial_y \Psi_L
+ c_I \delta_I 
\big( \partial_y \bar{\Psi}_L\big) \Psi_R
\Big\}.
\end{array}
\label{eq:lagrangian}
\end{equation}
A detailed discussion and interpretation of the 
coefficients $a_I, b_I, c_I$ can be found in Ref.~\cite{delAguila:2003bh}.
In the following we analyse a simple case, 
$b_I = c_I = 0$ and $a_I^L=0$, which will be sufficient to illustrate some
phenomenological implications of fermion BKT for popular models such
as Universal Extra Dimensions (UED) \cite{Appelquist:2000nn} and the
so-called Constrained Standard Model (CSM)~\cite{Barbieri:2000vh}. A
sum over the different irreducible representations of the Standard Model 
gauge group is
implicit. Each representation could have independent coefficients
$a_I^R$, but we assume that they are equal to minimize the number of
independent parameters. We choose $\Psi_R$ even
under the action of $Z_2$~\footnote{Note that we made the opposite 
choice in Ref.~\cite{delAguila:2003bh}.}; 
the representations of the 5D fields must
be chosen accordingly to reproduce the SM spectrum of zero modes.

\section{Kaluza Klein Spectrum}

Expanding in KK modes 
$
\Psi_{L,R}(x,y)=
\sum_n \frac{f^{L,R}_n(y)}{\sqrt{2\pi R}} \Psi^{(n)}_{L,R}(x)
$,
the 4D effective Lagrangian is diagonal in KK space with canonical
kinetic terms when the wave functions and 
masses satisfy the corresponding differential equations and
normalization conditions (see~\cite{delAguila:2003bh}). The solutions
are ($a_I\equiv a_I^R$)
%
%
a chiral massless zero mode:
\begin{equation}
f_0^R=\frac{1}{\sqrt{1+\frac{a_0+a_\pi}{2\pi R}}} \, ,
\label{eq:zeromode}
\end{equation}
and a tower of vector-like massive KK modes:
\begin{equation}
\begin{array}{rcl}
f_n^R&=&A_n [ \cos(m_n y)-\frac{a_0 m_n}{2} \sin(m_n |y|)],
\\
f_n^L&=& \partial_y f_n^R/m_n \,,
\end{array}
\label{eq:massivemodes}
\end{equation}
with the masses $m_n$ satisfying 
\begin{equation}
(4-a_0 a_\pi m_n^2) \tan(m_n \pi R)+2(a_0+a_\pi)m_n=0.
\label{eq:masses}
\end{equation}
Note that the solutions for the even functions $f_n^R$ (and the 
corresponding quadratic equations) are the same as for bosons
\cite{Carena:2002me} when only the coefficients $a_I^R$ are
nonvanishing. 

In figures 1 and 2 (left) we plot the masses for the first few 
KK modes as a function of the kinetic term at one brane 
$a_0 (a_{\pi} = 0)$ and at both branes $a_0 = a_{\pi}$, 
respectively. These plots coincide with 
the corresponding ones for gauge bosons~\cite{Carena:2002me}. 
In particular, for $a_{0,\pi}\gg R$ the lightest massive mode
approximates 
\begin{equation}
m_1^2 \sim 2\frac{a_0+a_\pi}{a_0 a_\pi \pi R}.
\end{equation}
In this limit this mode couples 
to the branes as the zero mode when $a_0 = a_\pi (\gg R)$ 
but decouples in the one brane case.

\begin{figure}[h]
\begin{center}
\epsfig{file=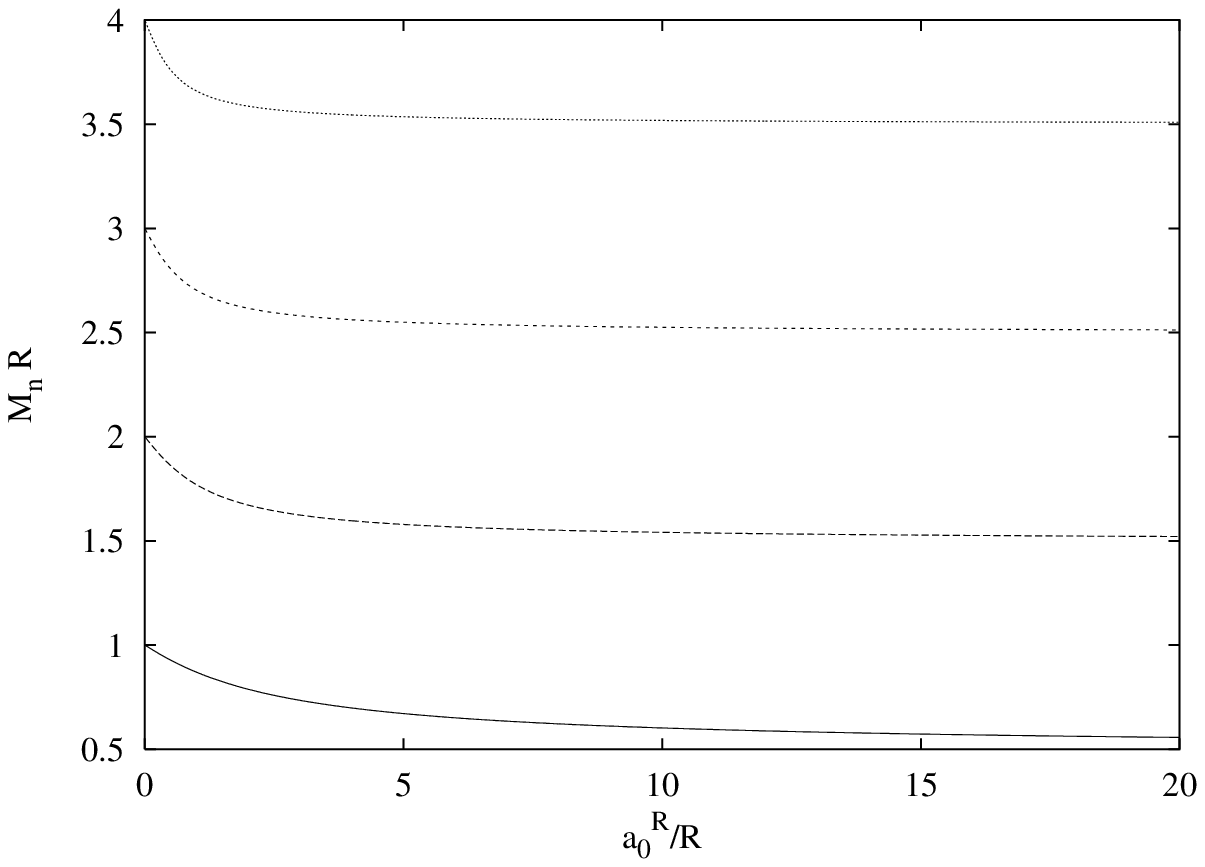,height=5cm}
\epsfig{file=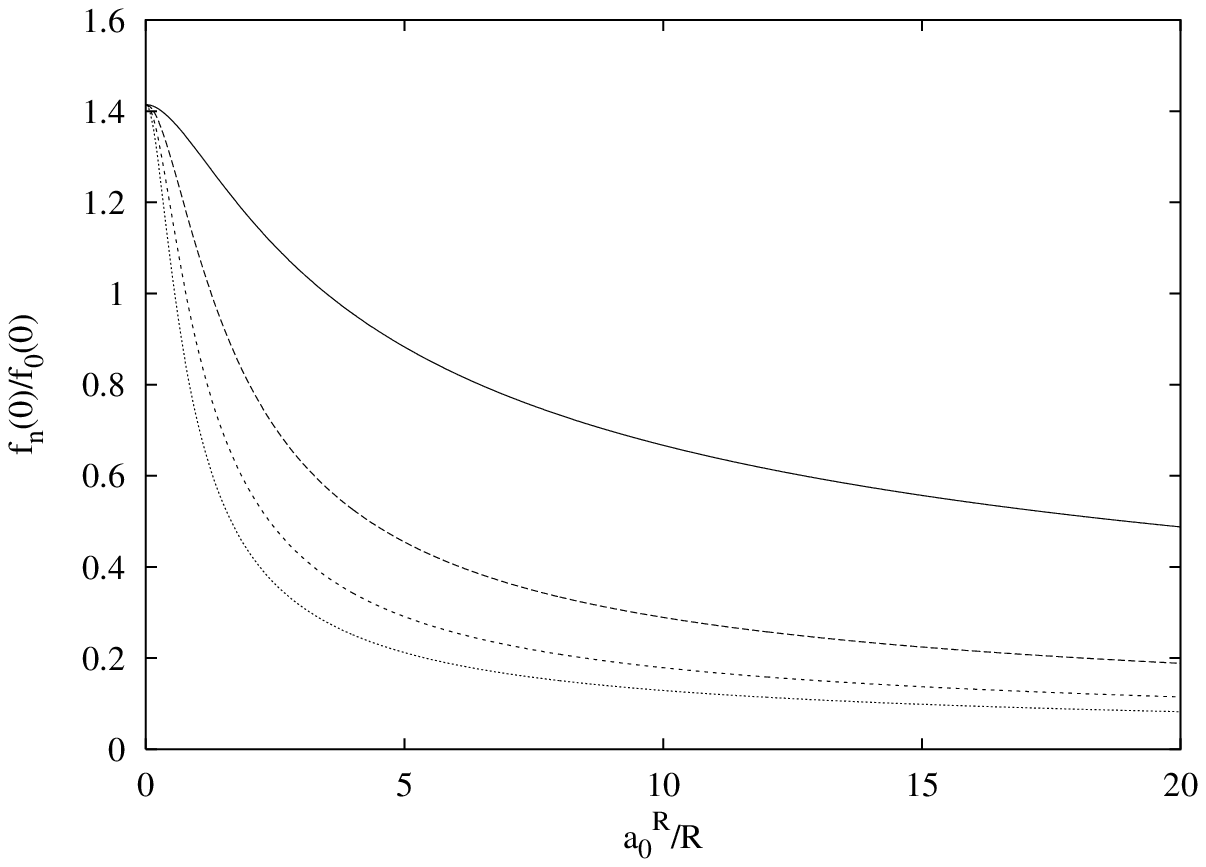,height=5cm} 
\caption{Masses (left) and couplings to the brane at $y=0$ in terms of
  the zero mode coupling (right) for the first few KK modes  
($n = 1,2,3,4$ from bottom to top (left) and from top to bottom
(right))
as a function of $a_0$, when $a_\pi=0$ .
\label{spectrum:onebrane}}
\end{center}
\end{figure}
\begin{figure}[h]
\begin{center}
\epsfig{file=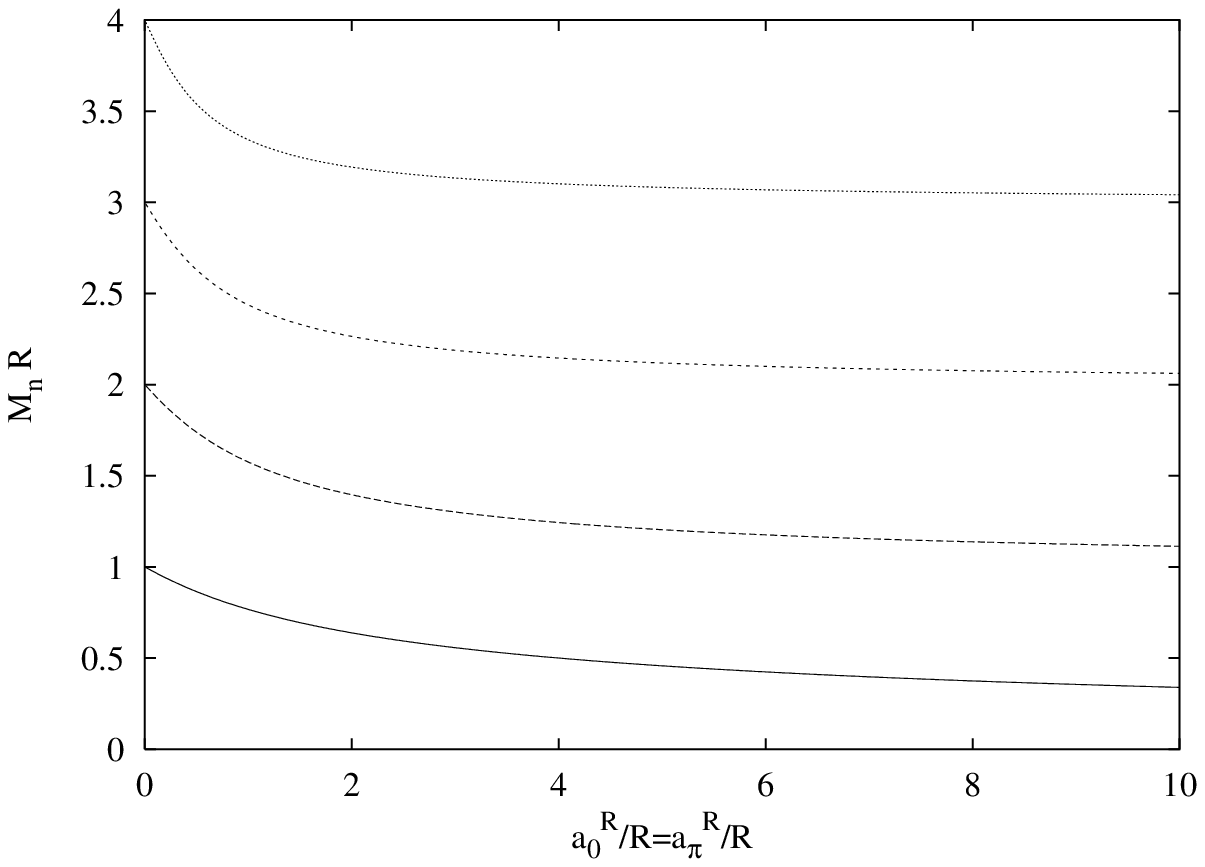,height=5cm}
\epsfig{file=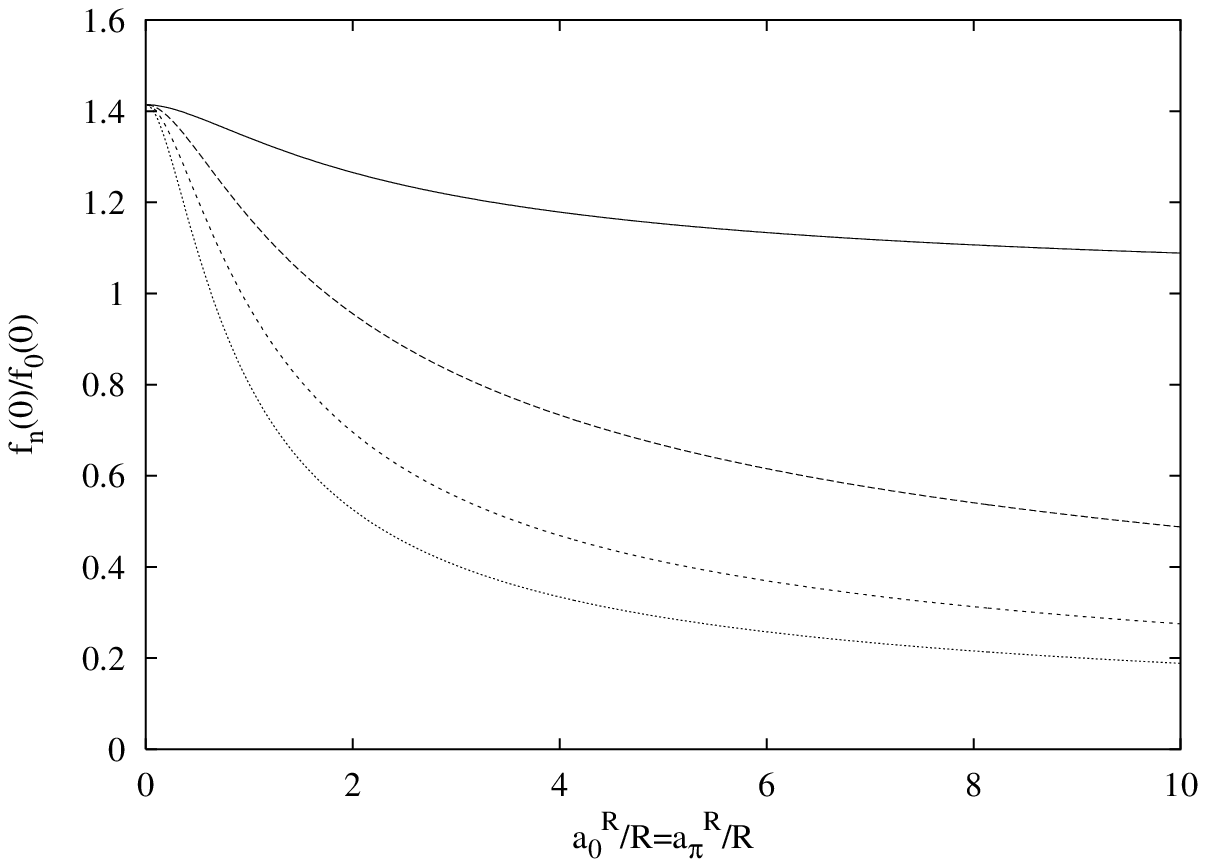,height=5cm} 
\caption{The same as in figure \ref{spectrum:onebrane} but with 
$a_0=a_\pi$.\label{spectrum:ued}}
\end{center}
\end{figure}

\section{Gauge Interactions}

The modification of the kinetic terms for fermions affects 
the full covariant derivative modifying the gauge couplings. 
The relevant part of the Lagrangian is 
\begin{equation}
{\mathcal L}=\int_{-\pi R}^{\pi R} \mathrm{d}y\; (1+ 
 a_0\delta_0 + a_\pi\delta_\pi )
\bar{\Psi}_R \mathrm{i} \gamma^\mu D_\mu \Psi_R+\ldots \,,
\end{equation}
leading to the effective couplings between fermion and 
gauge boson KK modes
\begin{equation}
g^{(mnr)}=\frac{g_5}{\sqrt{2\pi R}}\int_{-\pi R}^{\pi R} \mathrm{d}y\; 
(1+ a_0\delta_0 + a_\pi\delta_\pi )
\frac{ f_m^R f_n^R f_r^A }{2\pi R}. \label{effectivecoupling}
\end{equation}
The superscript $A$ refers to the gauge boson wave function. At low 
energy the integration of the KK tower of heavy gauge bosons 
gives rise to a four-fermion operator for the massless right-handed 
four-fermion field $\Psi_R$ (see Ref. 
\cite{Rizzo:1999br} for its definition), with coefficient
\begin{equation}
V=m_W^2\sum_{n>0} \frac{(g^{(00n)}/g^{(000)})^2}{ m_n^{A\ 2}}.
\end{equation}
From Eq.~\ref{effectivecoupling} we find
\begin{equation}
\frac{g^{(00n)}}{g^{(000)}}=
\frac{a_0 f_n^A(0)+a_\pi f_n^A(\pi R)}{2\pi R+a_0+a_\pi}.
\end{equation}
%
Taking into account the experimental limits on this coefficient (the
departure from the SM)~\cite{Rizzo:1999br},
\begin{equation}
V < \left
\{ \begin{array}{cl}
2.4 \times 10^{-3}, & \mbox{at LEP}, \\
1.25 \times 10^{-4}, & \mbox{at NLC},
\end{array}
\right.
\end{equation}
the exclusion region for the compactification scale $M_c\equiv1 / R$
can be 
estimated as a function of the brane kinetic parameters $a_{0,\pi}$ 
(in the absence of BKT for gauge bosons, $f_n^A(0) = (-1)^n f_n^A(\pi
R) = 1, m_n^A = n/R$).
In figure 3 we draw the corresponding forbidden regions.
\begin{figure}[h]
\begin{center}
\epsfig{file=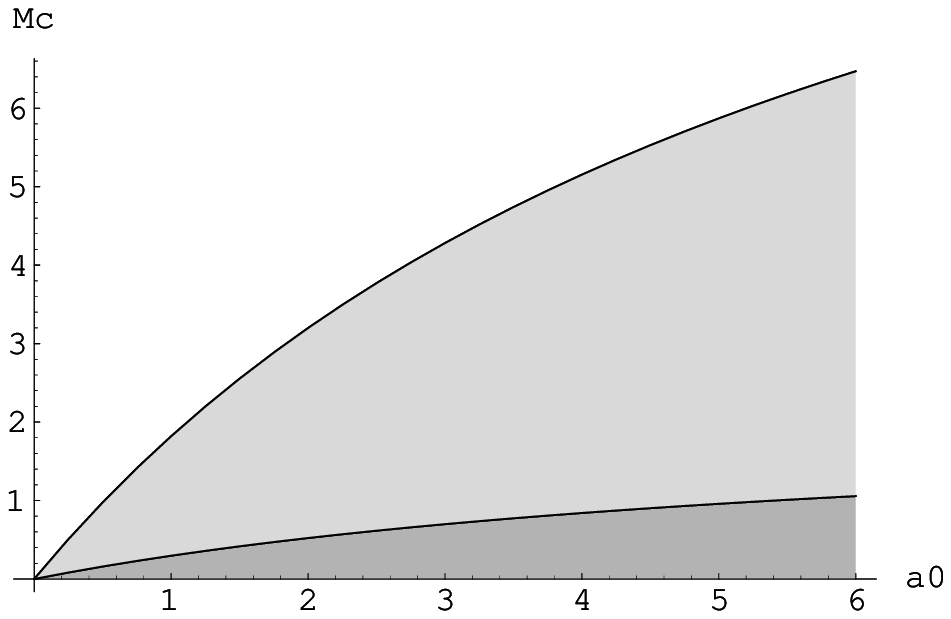,height=4cm} 
\epsfig{file=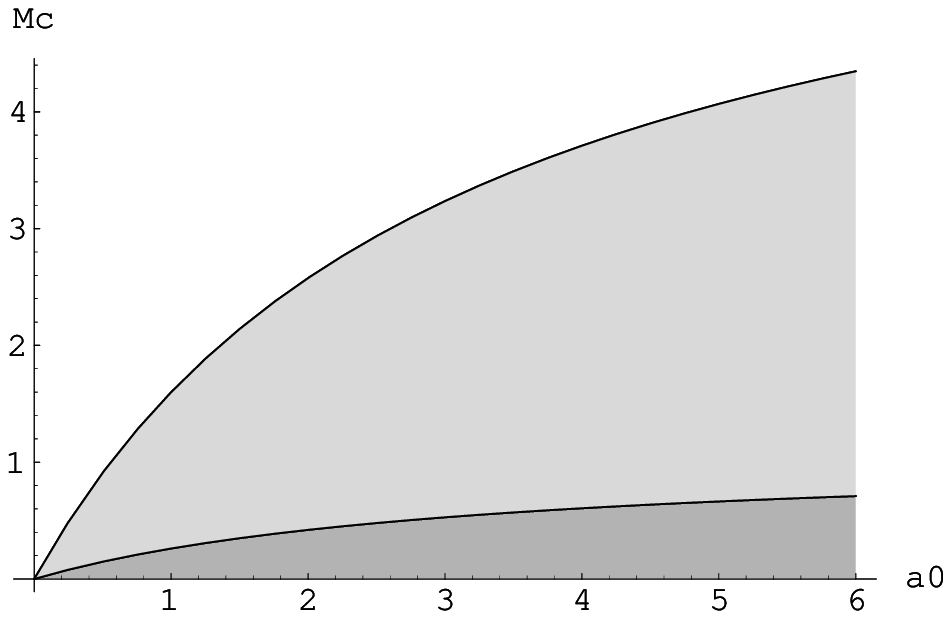,height=4cm} 
\caption{Forbidden regions from LEP data (dark) and expected reach of
NLC (shaded) in the case of BKT at one brane $a_{\pi} = 0$ (left) 
and BKT equal at both branes $a_{\pi} = a_0$ (right).\label{bounds:gauge}}
\end{center}
\end{figure}
It is worth noting that although the masses (wave functions) of the 
heavy KK fermions get more (less) 
reduced for BKT at just one brane than for 
BKT at both branes, as shown in figures 1 and 2, the excluded regions 
by LEP and NLC in figure 3 are larger in the former case. This is so 
because what determines the $V$ limits are the gauge boson wave 
functions, which alternate sign with $n$ at $y=\pi R$.
As can be read from the figure, in the case of UED
\cite{Appelquist:2000nn} (for which $a_0=a_\pi$) 
the bound coming from four-fermion contact interactions becomes more 
stringent for large BKT than the usually quoted one, $M_c > 300$ GeV,
coming from the $\rho$ parameter. The corrections to $\rho$ due to the
extra fermion mixing induced by BKT, to be discussed below, do not
improve the bounds in figure 3.

\section{Yukawa Couplings}

\begin{figure}[h]
\begin{center}
\epsfig{file=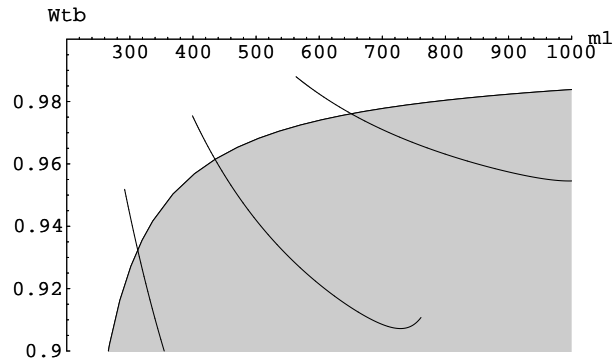,height=5.cm}
\caption{Value of the charged current top coupling as a function of
the mass of the first KK mode in the case of BKT and Yukawa couplinga 
at one and the same brane. The shaded region is forbidden by the
measurement of the $T$ parameter. The different lines correspond, from left 
to right, to $M_c=500,\;700,\; 1000$ GeV, and the BKT $a_0$ increases 
in each line from right to left from 0 to $20R$.
\label{constrained:plot}}
\end{center}
\end{figure}

For simplicity we will only consider boundary Yukawa couplings, which
are usually preferred to avoid consistency problems in supersymmetric
models or more severe flavour changing neutral current restrictions.
Then, the effective Yukawa couplings for the KK towers of 
fermions with given quantum numbers read (see, for 
example, Ref. \cite{delAguila:2001pu})
\begin{equation}
\lambda ^{(nm)}_{ij} = \frac {\lambda ^{(5)}_{ij}}{2\pi R}
f^L_n(0) f^R_m(0),
\end{equation}
with $i,j$ labelling the families. 
After electroweak symmetry breaking one obtains the standard
mass matrices for quarks and leptons 
\begin{equation}
M_{ij} = \lambda ^{(00)}_{ij} \frac {v}{\sqrt 2}.
\end{equation}
These mass matrices, however, are $3\times 3$ submatrices of 
the infinite mass matrix involving also the KK tower of 
vector-like fermions. Its diagonalisation gives corrections to the
masses of the observed quarks and leptons and to the CKM matrix, which 
is now nonunitary. The deviation from unitarity is produced by the
mixing with the heavy vector-like fermions~\cite{delAguila:2000rc}. 
This deviation can be shown to be proportional to the 
mass of the light fermion (and to the inverse of the heavy
one)~\cite{delAguila:2000kb}. A consequence of such scaling is that
this departure from the SM is likely to be seen only for the top
quark, whose charged gauge coupling to the bottom quark gets
corrected: 
\begin{equation}
W^L_{tb}\approx 1-\frac{1}{2}m_t^2\sum_{n>0} 
\Big(\frac{f^{t_R}_n(0)}{ f^{t_R}_0(0) m_n }\Big)^2.
\end{equation}
The effects of new vector-like fermions decouple with 
their masses, and hence with $M_c$ according to Eq. \ref{eq:masses}.
The most stringent limit from precision electroweak data results 
from the $T$ parameter \cite{Aguilar-Saavedra:2002kr}. 
For instance, in the absence of BKT Eqs. \ref{eq:zeromode} - 
\ref{eq:masses} imply
\begin{equation}
\frac{1}{R} > 2.5 \mbox{ TeV, for } W^L_{tb}=0.992.
\end{equation}
In the presence of BKT the masses and wave functions change, and so 
does this bound. 
Let us discuss the case of BKT and Yukawa couplings at one and the same
brane.  
In figure~\ref{constrained:plot} 
we draw the lines of constant $M_c$ and varying $a_0$ 
in the $W^L_{tb}-m_1$ plane, where $m_1$ is the mass of the lightest 
vector-like quark of charge $\frac{2}{3}$. $a_0$ grows along the lines 
from right to left, the initial (end) point corresponding to 
$a_0 = 0\ (20 R)$. The shaded region is forbidden by $T$. 
This means that models excluded without BKT 
can be recovered in their presence (moving along the lines to the left). 
This the case of the CSM~\cite{Barbieri:2000vh}, a supersymmetric
five-dimensional model compactified 
on the orbifold $S^1/(Z_2\times Z_2^\prime)$ with inverse radius
$1/R \approx 350$ GeV.
The $\mathcal{N}=2$ supersymmetry in the bulk forbids Yukawa couplings,
which have to be included at the boundaries, where the orbifold
action breaks supersymmetry down to $\mathcal{N}=1$. This model has
many interesting theoretical and phenomenological features, but
the mixing of the top quark with its KK excitations 
(with masses $m_n=2n M_c$ in this case, due to the extra orbifolding)
generates too large a $T$ parameter. This is apparent in
figure~\ref{constrained:plot}: remembering that 
$M_c$ in the figure is twice the compactification scale in this model,
we see that the right end of the line $M_c = 700$ GeV, corresponding 
to $1/R = 350$ GeV and no BKT, is excluded. However, moving 
to the left along the line, this model approaches the safe region, and
for large enough BKT it is reconciled with experiment 
~\footnote{This is not the case for $a_0=a_\pi$.}.

\section{Conclusions}

Models with extra dimensions typically have lower-dimensional defects on which
kinetic terms for the bulk fields can be localized. These BKT get
renormalised and therefore cannot be set to zero at all
scales. Moreover, it is natural that they be present already at tree
level. The BKT modify the wave functions and masses of
the KK modes of bulk fields, and thus their effective four-dimensional
couplings. We have discussed here the effects of BKT for bulk fermions,
for which relevant new effects appear both in gauge and Yukawa couplings.
The former are especially relevant in the case of 
UED~\cite{Appelquist:2000nn}, for which the bound on the
compactification scale coming from four-fermion 
contact interactions becomes the strictest one for BKT larger than
order $R$. The latter can be crucial to reconcile low scale models
with boundary Yukawa 
couplings, like the CSM \cite{Barbieri:2000vh}, 
with precision electroweak measurements. Signatures of this scenario 
would be the observation of a reduced $W^L_{tb}$ mixing and the 
observation of a relatively light quark singlet of charge $\frac{2}{3}$, 
as indicates figure~\ref{constrained:plot}. 
These signals are also compatible with five-dimensional models 
with multilocalized fermions \cite{delAguila:2001pu}.
We want to stress again, however, that these effects require large 
($\sim R$) BKT, which is not the case for BKT generated 
radiatively. But even small BKT can have observable 
phenomenological implications if they alter tree level {\it equalities}, 
as, for example, if they modify the mass degeneracies forbidding 
some decays~\cite{Cheng:2002iz}.

\section*{Acknowledgments}
This work has been supported in part by MCYT under contract 
FPA2000-1558, by Junta de Andaluc{\'\i}a group FQM 101, by 
the European Community's Human Potential Programme under 
contract HPRN-CT-2000-00149 Physics at Colliders, and by PPARC.

\end{document}